\documentclass[nocopyrightspace,10pt]{sigplanconf}
\usepackage{color}
\usepackage{alltt}
\usepackage{mathpartir}
\usepackage{hyperref}
\usepackage{amsmath, amssymb, amsthm}
\usepackage{stmaryrd}
\usepackage{xspace}
\usepackage{graphicx}
\usepackage{comment}
\usepackage{fouridx}
\usepackage{multirow}
\usepackage{tikz}
\usepackage{MnSymbol} 

\input{pyg-colors-ravi}

\def\parahead#1{\paragraph{\textbf{#1}.}}

\newcommand{\set}[1]{\{{#1}\}}
\newcommand{\seq}[1]{\overline{#1}}

\newcommand{\ie}{{\emph{i.e.,}\ }}
\newcommand{\eg}{{\emph{e.g.,}\ }}

\newcommand{\cf}{{\emph{cf.}\ }}

\newcommand{\javascript}{\text{JavaScript}}

\newcommand{\LamJS}{\ensuremath{\lambda_{JS}}}

\newcommand{\langBase}{\ensuremath{\textrm{D}}}
\newcommand{\dtypes}{\text{System}\ \langBase\xspace}
\newcommand{\dImp}{\text{System}\ !\ensuremath{\langBase}\xspace}
\newcommand{\dRef}{\dImp}

\newcommand{\djs}{\text{DJS}}

\newcommand{\tyBind}[2]{\ensuremath{{{#1}\!:\!{#2}}}}

\newcommand{\tyTupleTwo}[2]{(\ensuremath{{#1},\ {#2}})}
\newcommand{\tyTupleThree}[3]{(\ensuremath{{#1},\ {#2},\ {#3}})}

\newcommand{\expTupleTwo}[2]{(\ensuremath{{#1},\ {#2}})}
\newcommand{\expTupleThree}[3]{(\ensuremath{{#1},\ {#2},\ {#3}})}

\newcommand{\heapCatOp}{\oplus}

\newcommand{\anglesTwo}[2]{\langle{#1},{#2}\rangle}

\newcommand{\heapBind}[2]{\ensuremath{({#1}\mapsto{#2})}}
\newcommand{\heapBindPair}[3]{\ensuremath{({#1}\mapsto\anglesTwo{#2}{#3})}}

\newcommand{\heapCat}[2]{\ensuremath{{#1} \heapCatOp {#2}}}
\newcommand{\heapCatThr}[3]{\ensuremath{{#1} \heapCatOp \heapCat{#2}{#3}}}

\newcommand{\sameHeap}{\ensuremath{\mathit{same}}}

\newcommand{\curHeap}{\ensuremath{\mathit{cur}}}

\newcommand{\funBare}[2]{\ensuremath{\lambda{#1}.\ {#2}}}

\newcommand{\letinBare}[3]
   {\ensuremath{\texttt{let}\ {#1} = {#2}\ \texttt{in}\ {#3}}}

\newcommand{\ite}[3]
   {\ensuremath{\texttt{if}\ {#1}\ \texttt{then}\ {#2}\ \texttt{else}\ {#3}}}

\newcommand{\cast}[2]{\ensuremath{{#1}\ \texttt{as}\ {#2}}}
\newcommand{\dict}[1]{\ensuremath{\set{#1}}}
\newcommand{\emptydict}{\dict{}}

\newcommand{\singledict}[2]{{#1}\mapsto{#2}}
\newcommand{\dictextendOp}{\texttt{++}}
\newcommand{\dictextend}[3]{#1\ \dictextendOp\ \ensuremath{\singledict{#2}{#3}}}

\newcommand{\newref}[2]{\ensuremath{\texttt{ref}\ {#1}\ {#2}}}
\newcommand{\deref}[1]{\ensuremath{\texttt{deref}\ {#1}\ }}
\newcommand{\setref}[2]{\ensuremath{{#1}\mathtt\ {:=}\ {#2}}}

\newcommand{\freeze}[2]{\ensuremath{\texttt{freeze}\ {#1}\ \thawState\ {#2}}}
\newcommand{\thaw}[2]{\ensuremath{\texttt{thaw}\ {#1}\ {#2}}}

\newcommand{\appThree}[5]{\ensuremath{[{#1};{#2};{#3}]\ {#4}\ {#5}}}

\newcommand{\expLabel}[2]
  {\ensuremath{{\mathit{\labelNameX{#1}}}: {#2}}}
\newcommand{\expBreak}[2]
  {\ensuremath{\texttt{break}\ {\labelNameX{#1}}\ {{#2}}}}

\newcommand{\newObjThree}[3]{\ensuremath{\texttt{newobj}\ {#1}\ {#2}\ {#3}}}

\newcommand{\newAnn}{\ensuremath{\mathtt{ctor}}}

\newcommand{\nullCon}{{\ensuremath{\mathit{Null}}}}

\newcommand{\refCon}{\ensuremath{\mathit{Ref}}}
\newcommand{\arrayCon}{\ensuremath{\mathit{Arr}}}

\newcommand{\unfoldHeapHas}[4]
  {\ensuremath{\applyHelperOp{UnrollHas}{{#1},{#2},{#3},{#4}}}}
\newcommand{\unfoldHeapSel}[5]
  {\ensuremath{\applyHelperOp{UnrollSel}{{#1},{#2},{#3},{#4},{#5}}}}

\newcommand{\substOne}[2]{[{#2}/{#1}]}
\newcommand{\subst}[3]{{#1}\substOne{#2}{#3}}

\newcommand{\varVal}{v}
\newcommand{\varLogVal}{w}
\newcommand{\theV}{\nu}
\newcommand{\varTyp}{T}
\newcommand{\varScm}{S}
\newcommand{\varUnTyp}{U}

\newcommand{\varHeap}{H}

\newcommand{\varHeapBinding}{h}
\newcommand{\varHeapSigBinding}{\hat{h}}
\newcommand{\heapPair}[2]{(#1,#2)}
\newcommand{\varWorld}{W}
\newcommand{\tyvar}{A}
\newcommand{\tyvarB}{B}
\newcommand{\tyVar}{\tyvar}
\newcommand{\tyVarB}{\tyvarB}

\newcommand{\lprimP}{P}

\newcommand{\varFormOne}{p}
\newcommand{\varFormTwo}{q}

\newcommand{\varFormTwoSet}{Q}

\newcommand{\heapTyp}{\Sigma}
\newcommand{\empHeapTyp}{\emptyset}
\newcommand{\heapSig}{\hat{\Sigma}}
\newcommand{\weakHeap}{\Psi}
\newcommand{\locName}{\ell}
\newcommand{\weakLocName}{\locFrozen{\locName}}
\newcommand{\eitherLocName}{m}
\newcommand{\locConst}{a}

\newcommand{\varLoc}{L}

\newcommand{\varLocEither}{M}
\newcommand{\locFrozen}[1]{\tilde{#1}}
\newcommand{\frzn}{\ensuremath{\mathtt{frzn}}}
\newcommand{\thwd}[1]{\ensuremath\mathtt{thwd}\ {#1}}
\newcommand{\thawState}{\ensuremath{\theta}}
\newcommand{\locProto}[1]{\ensuremath{\dot{#1}}}

\newcommand{\runLoc}{\ensuremath{r}}
\newcommand{\varSubst}{\pi}
\newcommand{\emptySubst}{[]}

\newcommand{\labelName}{x}
\newcommand{\labelNameX}[1]{@{#1}}
\newcommand{\labelEnv}{\Omega}
\newcommand{\empLabelEnv}{\emptyset}

\newcommand{\defeq}{\stackrel{\circ}{=}}
\newcommand{\sepmidsep}{\ensuremath{\hspace{0.02in} | \hspace{0.02in}}}

\newcommand{\refTypShort}[1]{\ensuremath{\{#1\}}}
\newcommand{\refTyp}[1]{\ensuremath{\refTypShort{#1}}}
\newcommand{\refTypX}[2]{\ensuremath{\{ #1 \sepmidsep #2\}}}

\newcommand{\existsTyp}[3]{\ensuremath{\exists \tyBind{#1}{#2}.\ #3}}
\newcommand{\existsLoc}[2]{\ensuremath{\exists {#1}.\ #2}}

\newcommand{\typRef}[1]{\ensuremath{\refCon\miniSepOne #1}}

\newcommand{\instantiateName}{\ensuremath{\mathsf{TInst}}}
\newcommand{\instantiate}[3]
  {\ensuremath{\mathsf{\instantiateName(}{#1},{#2},{#3}\mathsf{)}}}

\newcommand{\vTrue}{\ensuremath{\mathtt{true}}}
\newcommand{\vFalse}{\ensuremath{\mathtt{false}}}

\newcommand{\vNull}{\ensuremath{\mathtt{null}}}
\newcommand{\vUndefined}{\ensuremath{\mathtt{undefined}}}

\newcommand{\vGet}{\mathtt{get}}

\newcommand{\ttfld}[1]{\ensuremath{\mathtt{``{#1}"}}}

\newcommand{\tyTop}{\ensuremath{\mathit{{Top}}}}

\newcommand{\tyStr}{\ensuremath{\mathit{Str}}}

\newcommand{\tyRecd}{\ensuremath{\mathit{Dict}}}
\newcommand{\tyDict}{\tyRecd}

\newcommand{\tyMaybeNull}[1]{\ensuremath{{#1}?}}

\newcommand{\formFalse}{\ensuremath{\mathit{false}}}
\newcommand{\formIteName}{\ensuremath{{\mathbf{ite}}}}
\newcommand{\formIte}[3]{\ensuremath{\formIteName\ {#1}\ {#2}\ {#3}}}

\newcommand{\truthy}[1]{\ensuremath{\mathit{truthy}(#1)}}
\newcommand{\falsy}[1]{\ensuremath{\mathit{falsy}(#1)}}

\newcommand{\formIffName}{\ensuremath{{\mathbf{iff}}}}
\newcommand{\formIff}[2]{\ensuremath{{#1}\ \formIffName\ {#2}}}

\newcommand{\logop}[1]{\ensuremath{\mathit{#1}}}

\newcommand{\objHasR}[4]
  {\ensuremath{\logop{ObjHas}({#1},{#2},{#3},{#4})}}

\newcommand{\domR}[1]{\ensuremath{\logop{dom}({#1})}}

\newcommand{\ttArr}{\ensuremath{{\rightarrow}}}

\newcommand{\utArrowDrefWorlds}[5]
  {\ensuremath{\forall[{#1};{#2};{#3}]\ {#4}\ \ttArr\ {#5}}}
\newcommand{\utArrowThree}[5]{\utArrowDrefWorlds{#1}{#2}{#3}{#4}{#5}}

\newcommand{\utRef}[1]{\ensuremath{\refCon\miniSepOne #1}}
\newcommand{\utArray}[1]{\ensuremath{\arrayCon({#1})}}

\newcommand{\ruleNameFig}[1]{\begin{scriptsize}[#1]\end{scriptsize}}
\newcommand{\ruleName}[1]{\textsc{\begin{normalsize}#1\end{normalsize}}}
\newcommand{\sep}{\hspace{0.06in}}
\newcommand{\sepPremise}{\hspace{0.20in}}
\newcommand{\hsepRule}{\hspace{0.20in}}
\newcommand{\vsepRule}{\vspace{0.08in}}
\newcommand{\miniSepOne}{\hspace{0.01in}}
\newcommand{\miniSepTwo}{\hspace{0.02in}}

\newcommand{\startRow}[3]{{#1} & \sep {#2} & \sep {#3} \sep &}

\newcommand{\spaceItem}{\sep\mid\sep}
\newcommand{\spaceItemSmall}{\miniSepTwo\mid\miniSepTwo}
\newcommand{\spaceCategory}{\\[3pt]}

\newcommand{\gap}{\vspace{0.05in}}
\newcommand{\dsRuleNameFig}[1]{\hfill\ruleNameFig{\textsc{#1}}}

\newcommand{\helperOp}[1]{\ensuremath{\mathsf{#1}}}
\newcommand{\applyHelperOp}[2]{\ensuremath{\helperOp{#1}\mathsf{(}{#2}\mathsf{)}}}

\newcommand{\hInst}[3]{\ensuremath{\applyHelperOp{HInst}{{#1},{#2},{#3}}}}
\newcommand{\lInst}[3]{\ensuremath{\applyHelperOp{LInst}{{#1},{#2},{#3}}}}
\newcommand{\tInst}[3]{\ensuremath{\applyHelperOp{TInst}{{#1},{#2},{#3}}}}

\newcommand{\cnfName}{\ensuremath{\mathsf{CNF}}}
\newcommand{\cnf}[1]{\ensuremath{\mathsf{\cnfName(}#1\mathsf{)}}}

\newcommand{\typConst}[1]{\ensuremath{ty(#1)}}

\newcommand{\valid}[1]{\ensuremath{\mathsf{Valid(}#1\mathsf{)}}}
\newcommand{\embed}[1]{\ensuremath{{\llbracket #1 \rrbracket}}}

\newcommand{\freshen}[1]{\ensuremath{\applyHelperOp{Freshen}{#1}}}

\newcommand{\typePredName}{\ensuremath{::}}
\newcommand{\hasTyp}[2]{\ensuremath{{#1}\typePredName{#2}}}

\newcommand{\heapHas}[3]{\ensuremath{\logop{HeapHas}({#1},{#2},{#3})}}

\newcommand{\expand}[1]{\ensuremath{\applyHelperOp{Unroll}{#1}}}

\newcommand{\iteJoin}[3]{\ensuremath{\applyHelperOp{Join}{#1,#2,#3}}}
\newcommand{\iteJoinTypes}[3]{\ensuremath{\applyHelperOp{JoinTypes}{#1,#2,#3}}}
\newcommand{\iteJoinHeaps}[3]{\ensuremath{\applyHelperOp{JoinHeaps}{#1,#2,#3}}}

\newcommand{\symSub}{\sqsubseteq}
\newcommand{\symSynSub}{<:}

\newcommand{\cnfClause}[2]{\ensuremath{#1 \Rightarrow #2}}

\newcommand{\relImplForm}[2]{\ensuremath{#1\Rightarrow {#2}}}

\newcommand{\world}[2]{\ensuremath{{#1}\miniSepOne /\miniSepOne {#2}}}
\newcommand{\triWorld}[3]{\world{\tyBind{#1}{#2}}{#3}}

\newcommand{\relTypVal}[5]
  {\ensuremath{#1;\ #2 \vdash_{#5}
   #3\hspace{0.02in}::\hspace{0.02in}#4}}

\newcommand{\relTypExp}[6]
  {\ensuremath{#1;\ #2;\ \labelEnv \vdash_{#6}
   #3\hspace{0.02in}::\hspace{0.02in}\world{#4}{#5}}}
\newcommand{\relTypExpLabel}[7]
  {\ensuremath{#1;\ #2;\ #3\vdash_{#7} #4
   \hspace{0.02in}::\hspace{0.02in}\world{#5}{#6}}}

\newcommand{\relSub}[4]
  {\ensuremath{#1\vdash_{#4} #2\hspace{0.02in}\symSub \hspace{0.02in}#3}}
\newcommand{\relHeapSub}[5]
  {\ensuremath{#1\vdash_{#5} #2\hspace{0.02in}\symSub \hspace{0.02in}#3;\ #4}}

\newcommand{\relHeapSat}[5]
  {\ensuremath{#1\vdash_{#5} #2\hspace{0.02in}\models \hspace{0.02in}#3;\ #4}}

\newcommand{\relWorldSat}[5]
  {\ensuremath{#1\vdash_{#5} #2\hspace{0.02in}\models \hspace{0.02in}#3;\ #4}}

\newcommand{\relMatch}[3]{\ensuremath{#1 \sim #2 ;\ #3}}

\newcommand{\heapEqSym}{\equiv}
\newcommand{\heapEq}[2]{\ensuremath{{#1}\heapEqSym{#2}}}

\newcommand{\relSynSub}[4]{\ensuremath{#1\vdash_{#4}{#2}\symSynSub{#3}}}

\newcommand{\relWf}[2]{\ensuremath{#1\vdash #2}}

\newcommand{\relDescription}[1]{\ensuremath{\textrm{\textbf{#1}}}}
\newcommand{\judgementHeadNoBox}[2]{\ensuremath{\relDescription{#1}\hfill{#2}}}
\newcommand{\judgementHead}[2]{\judgementHeadNoBox{#1}{\fbox{#2}}}

\newcommand{\judgementHeadThree}[3]
  {\ensuremath{\relDescription{#1}\ \textrm{{#2}}\hfill\fbox{#3}}}

\newcommand{\djsFont}[1]{\ensuremath{\mathtt{#1}}}
\newcommand{\desugarWrap}[1]{\ensuremath{\llangle\ {#1}\ \rrangle}}
\newcommand{\desugarWrapSlim}[1]{\ensuremath{\llangle{#1} \rrangle}}
\newcommand{\desugarSym}{\ensuremath{=}}

\newcommand{\dsLHS}[1]{\ensuremath{\desugarWrap{\djsFont{#1}}}}
\newcommand{\ds}[1]{\ensuremath{\desugarWrapSlim{\djsFont{#1}}}}
\newcommand{\dsVar}[1]{\ensuremath{\_{#1}}}
\newcommand{\dsAnn}[1]{\ensuremath{/*:{#1}*/}}

\newcommand{\dsFuncCode}[1]
  {\ensuremath{\ullcorner{\djsFont{#1}}\ulrcorner}}
\newcommand{\dsGetProto}[1]{\ensuremath{\mathit{pro}(#1)}}

\renewcommand{\theFancyVerbLine}
  {\sffamily \textcolor[rgb]{0.1,0.1,0.1}
    {\scriptsize \oldstylenums{\arabic{FancyVerbLine}}}}

\definecolor{bg_js}{rgb}{0.93,0.93,0.93}
\definecolor{bg_dref}{rgb}{1,1,1}
\definecolor{line_white}{rgb}{0,0,0}

\newcommand{\beginMintedJs}[1]
  {\vspace{0.10in}\begin{minted}
    [numbersep=5pt,bgcolor=bg_js,fontfamily=tt,fontsize=\small,frame=single,mathescape,{#1}]{js}}
\newcommand{\beginMintedDref}[1]
  {\vspace{0.10in}\begin{minted}
    [numbersep=5pt,bgcolor=bg_dref,fontfamily=tt,fontsize=\small,frame=single,mathescape,{#1}]{ocaml}}

\newcommand{\beginMintedJsBigger}[1]
  {\vspace{0.10in}\begin{minted}
    [bgcolor=bg_js,fontfamily=tt,frame=single,mathescape,{#1}]{js}}

\newcommand{\figureFontSize}{\small}
\newcommand{\equationSizeBegin}{\small}
\newcommand{\equationSizeEnd}{\normalsize}

\newif\ifUseIndices
\newcommand{\maybeN}{\ifUseIndices n \else \fi}

\newif\ifLongVersion

\newcommand{\refSecsyntaxandsemantics}{\S 3}
\newcommand{\refSectypechecking}{\S 4}
\newcommand{\refSectypecheckingexp}{\S 4.4}
\newcommand{\refSecdesugaring}{\S 5}
\newcommand{\refSecoverviewcollections}{\S 2.7}
\newcommand{\refFigsyntaxdref}{Figure 7}
\newcommand{\refFigdectypingvaluesselected}{Figure 9}
\newcommand{\refFigdecsubtyping}{Figure 8}
\newcommand{\refFigdectypingexpselected}{Figure 10}
\newcommand{\refFigdesugaring}{Figure 12}

\title{Dependent Types for JavaScript --- Appendix\thanks{
This report supplements our OOPSLA 2012 paper~\cite{DjsOOPSLA12}.}}

\authorinfo
  {Ravi Chugh}
  {University of California, San Diego}
  {rchugh@cs.ucsd.edu}

\authorinfo
  {David Herman}
  {Mozilla Research}
  {dherman@mozilla.com}

\authorinfo
  {Ranjit Jhala}
  {University of California, San Diego}
  {jhala@cs.ucsd.edu}

\begin{document}

\maketitle

\appendix

\setcounter{figure}{13}

\section{Additional Definitions}
\label{sec:appendix-a}

We presented many parts of $\dRef$ and $\djs$ in
\refSecsyntaxandsemantics, \refSectypechecking, and \refSecdesugaring.
In this appendix, we consider some details that did not
fit in that presentation, as well as our treatment of
\emph{break} and \emph{label} expressions to facilitate the
desugaring of control operators in $\javascript$.
Then, in \autoref{sec:appendix-extensions}, we outline how
to extend $\dRef$ to support better \emph{location polymorphism}.

\subsection{Syntax}

In addition to the expression syntax in \refFigsyntaxdref, $\dRef$
includes the following forms:
$$
\figureFontSize
\begin{array}{rrcll}
\startRow{\textbf{}}{e}{::=}
  {\cdots}
  \spaceItem
  {\expLabel{\labelName}{e}}
  \spaceItem
  {\expBreak{\labelName}{\varVal}}
\end{array}
$$
An expression $\expLabel{x}{e}$ labels the enclosed expression,
and a break expression $\expBreak{\labelName}{\varVal}$ terminates
execution of the innermost expression labeled $\labelNameX{\labelName}$
within the function currently being evaluated and produces the result
$\varVal$.
If no such labeled expression is found, evaluation becomes stuck.
Label and break expressions are included to translate the control
flow operations of $\djs$.

To analyze label and break expressions, the expression typing relation
uses a \emph{label environment} $\labelEnv$ (in addition to type and
heap environments), where each binding records the world that the
expression labeled $\labelNameX{\labelName}$ is expected to satisfy.
$$
\figureFontSize
\begin{array}{rrcll}
\startRow{\textbf{}}{\labelEnv}{::=}
  {\emptyset}
  \spaceItem
  {\labelEnv,\tyBind{\labelNameX{\labelName}}
                                {(\world{\varTyp}{\heapSig})}}
\end{array}
$$

\subsection{Well-Formedness}

{
\begin{figure}[t]
\figureFontSize
\centering


\judgementHead{Well-Formed Types}{$\relWf{\Gamma}{\varTyp}$}

\vsepRule

$\inferrule*
  {\relWf{\Gamma,\tyBind{x}{\tyTop}}{\varFormOne}}
  {\relWf{\Gamma}{\refTypX{x}{\varFormOne}}}$
\hsepRule
$\inferrule*
  {\relWf{\Gamma}{\varTyp} \sepPremise
   \relWf{\Gamma, \tyBind{x}{\tyTop}}{\varScm}}
  {\relWf{\Gamma}{\existsTyp{x}{\varTyp}{\varScm}}}$

\vsepRule



\judgementHeadThree
   {Well-Formed Formulas}{(selected rules)}
   {$\relWf{\Gamma}{\varFormOne}$}

\vsepRule

$\inferrule*
  {\relWf{\Gamma}{\varLogVal} \sepPremise
   \relWf{\Gamma}{\varUnTyp}}
  {\relWf{\Gamma}{\hasTyp{\varLogVal}{\varUnTyp}}}$
\sep\sep
$\inferrule*
  {\relWf{\Gamma}{\seq{\varLogVal}}}
  {\relWf{\Gamma}{\lprimP(\seq{\varLogVal})}}$
\sep\sep
%
$\inferrule*
  {\varHeap\in\Gamma \sepPremise
   \relWf{\Gamma}{\locName} \sepPremise
   \relWf{\Gamma}{\varLogVal}}
  {\relWf{\Gamma}{\heapHas{\varHeap}{\locName}{\varLogVal}}}$

\vsepRule




\judgementHead{Well-Formed Locations}{$\relWf{\Gamma}{\locName}$}

\vsepRule

$\inferrule*
  { }
  {\relWf{\Gamma}{\locConst}}$
\hsepRule
$\inferrule*
  { }
  {\relWf{\Gamma}{\locFrozen{\locConst}}}$
\hsepRule
$\inferrule*
  {\varLoc\in\Gamma}
  {\relWf{\Gamma}{\varLoc}}$
%

\vsepRule


\judgementHead{Well-Formed Type Terms}{$\relWf{\Gamma}{\varUnTyp}$}

\vsepRule

$\inferrule*
  {
   \Gamma_1 =
     \Gamma, \seq{\tyVar}, \seq{\varLoc}, \seq{\varHeap} \sepPremise
   \relWf{\Gamma_1}{\varWorld_1} \sepPremise
   \relWf{\Gamma_1, \applyHelperOp{Binders}{\varWorld_1}}{\varWorld_2}
  }
  {\relWf{\Gamma}
      {\utArrowDrefWorlds
          {\seq{\tyVar}}{\seq{\varLoc}}{\seq{\varHeap}}
          {\varWorld_1}{\varWorld_2}}}$

\vsepRule

$\inferrule*
  {\tyVar\in\Gamma}
  {\relWf{\Gamma}{\tyVar}}$
\hsepRule
$\inferrule*
  {\relWf{\Gamma}{\varTyp}}
  {\relWf{\Gamma}{\utArray{\varTyp}}}$
\hsepRule
$\inferrule*
  {\relWf{\Gamma}{\eitherLocName}}
  {\relWf{\Gamma}{\utRef{\eitherLocName}}}$
%
\hsepRule
$\inferrule*
  { }
  {\relWf{\Gamma}{\nullCon}}$

\vsepRule



\judgementHead{Well-Formed Worlds}{$\relWf{\Gamma}{\varWorld}$}

\vsepRule

$\inferrule*
  {\relWf{\Gamma}{\varScm} \sepPremise
   \relWf{\Gamma, \tyBind{x}{\tyTop}}{\heapSig}}
  {\relWf{\Gamma}{\world{\tyBind{x}{\varScm}}{\heapSig}}}$

 \vsepRule


\judgementHeadNoBox{Well-Formed Heaps and Heap Bindings}
                   {\fbox{$\relWf{\Gamma}{\heapSig}$}\
                    \fbox{$\relWf{\Gamma}{\varHeapSigBinding}$}}

\vsepRule

$\inferrule*
  {
   \varHeap \in \Gamma \sepPremise
   \relWf{\Gamma,\applyHelperOp{Binders}{\varHeapSigBinding}}
         {\varHeapSigBinding} \sepPremise
   \textrm{no duplicate locations in } \varHeapSigBinding
  }
  {\relWf{\Gamma}{\heapPair{\varHeap}{\varHeapSigBinding}}}$

\vsepRule


$\inferrule*
  {\relWf{\Gamma}{\locName} \sepPremise
   \relWf{\Gamma}{\varTyp}}
  {\relWf{\Gamma}{\heapBind{\locName}{\tyBind{x}{\varTyp}}}}$
\sep
$\inferrule*
  {\relWf{\Gamma}{\locName} \sepPremise
   \relWf{\Gamma}{\varTyp} \sepPremise
   \relWf{\Gamma}{\locName'}}
  {\relWf{\Gamma}{\heapBindPair{\locName}{\tyBind{x}{\varTyp}}{\locName'}}}$
\sep
$\inferrule*
  { }
  {\relWf{\Gamma}{\empHeapTyp}}$

\vsepRule

$\inferrule*
  {\relWf{\Gamma}{\weakLocName}}
  {\relWf{\Gamma}{\heapBind{\weakLocName}{\frzn}}}$
\sep
$\inferrule*
  {\relWf{\Gamma}{\weakLocName} \sepPremise
   \relWf{\Gamma}{\locName}}
  {\relWf{\Gamma}{\heapBind{\weakLocName}{\thwd{\locName}}}}$
\sep
$\inferrule*
  {\relWf{\Gamma}{\varHeapSigBinding_1} \sepPremise
   \relWf{\Gamma}{\varHeapSigBinding_2}}
  {\relWf{\Gamma}{\heapCat{\varHeapSigBinding_1}{\varHeapSigBinding_2}}}$

\vsepRule

\caption{Well-formedness for \dImp}
\label{fig:well-formedness}
\end{figure}

}

The well-formedness relations, defined in \autoref{fig:well-formedness},
are largely straightforward.
We use the procedure $\helperOp{Binders}$ to collect all of the binders
in a world or heap.

\subsection{Subtyping}

{

\begin{figure}[t]
\figureFontSize
\centering

\judgementHeadThree{Subtyping}
   {(see \refFigdecsubtyping)}
   {$\relSub{\Gamma}{\varTyp_1}{\varTyp_2}{\maybeN}$}

\vsepRule

\judgementHeadThree{Syntactic Subtyping}
   {(extends \refFigdecsubtyping)}
   {$\relSynSub{\Gamma}{\varUnTyp_1}{\varUnTyp_2}{\maybeN}$}

\vsepRule

$\inferrule*[lab=\ruleNameFig{U-Arrow}]
  {
   \relHeapSub{\Gamma}
              {\varWorld_{21}}{\varWorld_{11}} {\varSubst}{\maybeN} \sepPremise
   \relHeapSub{\Gamma,\embed{\varWorld_{21}}}
              {\varSubst\varWorld_{12}}{\varWorld_{22}}{\varSubst'}{\maybeN}
  }
  {\relSynSub{\Gamma}
      {\utArrowDrefWorlds{\seq{\tyVar}}{\seq{\varLoc}}{\seq{\varHeap}}
        {\varWorld_{11}}{\varWorld_{12}}}
      {\utArrowDrefWorlds{\seq{\tyVar}}{\seq{\varLoc}}{\seq{\varHeap}}
        {\varWorld_{21}}{\varWorld_{22}}}
      {\maybeN}}$

\vsepRule


\judgementHead{Implication}
              {$\relImplForm{\Gamma}{\varFormOne}$}

\vsepRule

$\inferrule*[right=\ruleNameFig{I-Cnf}]
  {\cnf{\varFormOne} =
     \wedge_i\ (\cnfClause{\varFormOne_i}{\varFormTwoSet_i}) \sepPremise
   \forall i.\ \exists q \in \varFormTwoSet_i.\ 
     \relImplForm{\Gamma, \varFormOne_i}{\varFormTwo}}
  {\relImplForm{\Gamma}{\varFormOne}}
$

\vsepRule

$\inferrule*[lab=\ruleNameFig{I-Valid}]
  {\valid{\embed{\Gamma}\Rightarrow\varFormOne}}
  {\relImplForm{\Gamma}{\varFormOne}}
$
\hsepRule
$\inferrule*[lab=\ruleNameFig{I-HasTyp}]
  {\valid{\embed{\Gamma}\Rightarrow\hasTyp{\varLogVal}{\varUnTyp'}}
      \sepPremise
   \relSynSub{\Gamma}{\varUnTyp'}{\varUnTyp}{\maybeN}}
  {\relImplForm{\Gamma}{\hasTyp{\varLogVal}{\varUnTyp}}}
$

\vsepRule


\judgementHead{World Subtyping}
              {$\relHeapSub{\Gamma}{\varWorld_1}{\varWorld_2}{\varSubst}{\maybeN}$}

\vsepRule


$\inferrule*
  {
   \relSub{\Gamma}{\varTyp_1}{\varTyp_2}{\maybeN} \sepPremise
   \relMatch{\varHeapSigBinding_1}{\varHeapSigBinding_2}{\varSubst} \\\\
   \varSubst' = \varSubst\substOne{x_2}{x_1} \sepPremise
   \relImplForm{\Gamma, \embed{\varHeapSigBinding_1}}
               {\varSubst\embed{\varHeapSigBinding_2}}
  }
  {\relSub{\Gamma}
      {\world{\tyBind{x_1}{\varTyp_1}}{\heapPair{\varHeap}{\varHeapSigBinding_1}}}
      {\world{\tyBind{x_2}{\varTyp_2}}{\heapPair{\varHeap}{\varHeapSigBinding_2}}}
      {\maybeN}}
$

\vsepRule



\judgementHead{Heap Matching}
              {$\relMatch{\varHeapSigBinding_1}
                         {\varHeapSigBinding_2}
                         {\varSubst}$}

\vsepRule

$\inferrule*
   { }
   {\relMatch{\empHeapTyp}{\empHeapTyp}{[]}}$
%
\sep
$\inferrule*
   {\heapEq{\varHeapSigBinding_1}{\varHeapSigBinding_1'} \sepPremise
    \heapEq{\varHeapSigBinding_2}{\varHeapSigBinding_2'} \sepPremise
    \relMatch{\varHeapSigBinding_1'}{\varHeapSigBinding_2'}{\varSubst}}
   {\relMatch{\varHeapSigBinding_1}{\varHeapSigBinding_2}{\varSubst}}$

\vsepRule

$\inferrule*
   {\relMatch{\varHeapSigBinding_1}{\varHeapSigBinding_2}{\varSubst}}
   {\relMatch{\heapCat{\varHeapSigBinding_1}{\heapBind{\weakLocName}{\thawState}}}
             {\heapCat{\varHeapSigBinding_2}{\heapBind{\weakLocName}{\thawState}}}
             {\varSubst}}$

\vsepRule

$\inferrule*
   {\relMatch{\varHeapSigBinding_1}{\varHeapSigBinding_2}{\varSubst}}
   {\relMatch{\heapCat{\varHeapSigBinding_1}{\heapBind{\locName}{\tyBind{x}{\varTyp}}}}
             {\heapCat{\varHeapSigBinding_2}{\heapBind{\locName}{\tyBind{y}{\varScm}}}}
             {\varSubst\substOne{y}{x}}}$

\vsepRule

$\inferrule*
   {\relMatch{\varHeapSigBinding_1}{\varHeapSigBinding_2}{\varSubst}}
   {\relMatch{\heapCat{\varHeapSigBinding_1}
                      {\heapBindPair{\locName}{\tyBind{x}{\varTyp}}{\locName'}}}
             {\heapCat{\varHeapSigBinding_2}
                      {\heapBindPair{\locName}{\tyBind{y}{\varScm}}{\locName'}}}
             {\varSubst\substOne{y}{x}}}$

\vsepRule

\caption{Subtyping for \dImp}
\label{fig:dec-subtyping-heaps}

\end{figure}

}

\autoref{fig:dec-subtyping-heaps} presents more of the subtyping relations.

\parahead{Implication}

As in $\dtypes$, subtyping on refinement types
reduces to implication of refinement formulas, which are discharged by a
combination of uninterpreted, first-order reasoning and syntactic subtyping.
If the SMT solver alone cannot discharge an implication
obligation ($\ruleName{I-Valid}$), the formula is rearranged into
conjunctive normal form ($\ruleName{I-Cnf}$), and goals of the form
$\hasTyp{\varLogVal}{\varUnTyp}$ are discharged by a combination of
uninterpreted reasoning and syntactic subtyping ($\ruleName{I-ImpSyn}$).

We write $\embed{\varTyp}$ for the \emph{embedding} of a type as a
formula, a straightforward definition \cite{NestedPOPL12} that lifts to
environments $\embed{\Gamma}$, heap bindings $\embed{\varHeapSigBinding}$,
heaps $\embed{\heapSig}$, and worlds $\embed{\varWorld}$.
Because heap binders may refer to each other in any order
(recall that a heap can be thought of as a dependent tuple, where
each component is named with a binder), the embedding of a heap starts
by inserting dummy bindings so that all binders in scope for the type of each
heap binding.
For example:
\equationSizeBegin
\begin{align*}
\heapSig_0 &\defeq
   \heapPair{\varHeap_0}
            {\heapCat{\heapBind{\locName_1}{\tyBind{x}{\varTyp_1}}}
                     {\heapBind{\locName_2}{\tyBind{y}{\varTyp_2}}}} \\
\embed{\heapSig_0} & =
    \embed{\tyBind{x}{\tyTop}},
    \embed{\tyBind{y}{\tyTop}},
    \embed{\varTyp_1}(x),
    \embed{\varTyp_2}(y)
\end{align*}
\equationSizeEnd

\parahead{Syntactic Subtyping}

The $\ruleName{U-Arrow}$ rule for function types is familiar,
treating input worlds contravariantly and output worlds covariantly.

\parahead{Worlds}

In order to check world subtyping, the judgement
$\relSub{\Gamma}
        {\world{\tyBind{x_1}{\varTyp_1}}{\heapPair{\varHeap}{\varHeapSigBinding_1}}}
        {\world{\tyBind{x_2}{\varTyp_2}}{\heapPair{\varHeap}{\varHeapSigBinding_2}}}
        {\maybeN}$
checks that $\varTyp_1$ is a subtype of $\varTyp_2$ and
that the heaps agree on the ``deep" part $\varHeap$.
Then, it checks that the structure of the
``shallow" parts match --- using a heap matching relation that uses
a $\heapEqSym$ operator (not shown) that permutes bindings as necessary ---
and creates a substitution $\varSubst$ of binders
from $\varHeapSigBinding_2$ to $\varHeapSigBinding_1$.
Finally, the heap bindings, which can be thought of as dependent tuples,
are embedded as formulas and checked by implication.

\subsection{Value Typing}


\begin{figure}[t]
\figureFontSize
\centering

\judgementHead{Value Typing}
  {$\relTypVal{\Gamma}{\heapTyp}{\varVal}{\varTyp}{\maybeN}$}

\vsepRule

$\inferrule*[lab=\ruleNameFig{T-Const}]
  { }
  {\relTypVal{\Gamma}{\heapTyp}{c}{\typConst{c}}{\maybeN}}
$
\hsepRule
$\inferrule*[lab=\ruleNameFig{T-Var}]
  {\Gamma(x)=\varScm}
  {\relTypVal{\Gamma}{\heapTyp}{x}{\refTypX{y}{y=x}}{\maybeN}}
$

\vsepRule

$\inferrule*[lab=\ruleNameFig{T-Loc}]
  {\applyHelperOp{StaticLoc}{\runLoc} = \eitherLocName \sepPremise
   \eitherLocName \in \domR{\heapTyp}}
  {\relTypVal{\Gamma}{\heapTyp}{\runLoc}
     {\refTypX{x}{x=\runLoc \wedge \hasTyp{x}{\utRef{\eitherLocName}}}}{\maybeN}}
$

\vsepRule

$\inferrule*[lab=\ruleNameFig{T-Extend}]
  {\relTypVal{\Gamma}{\heapTyp}
             {\expTupleThree{\varVal_1}{\varVal_2}{\varVal_3}}
             {\tyTupleThree{\tyRecd}{\tyStr}{\varTyp}}{\maybeN}}
  {\relTypVal{\Gamma}{\heapTyp}
             {\dictextend{\varVal_1}{\varVal_2}{\varVal_3}}
             {\refTypX{x}{x=\dictextend{\varVal_1}{\varVal_2}{\varVal_3}}}
             {\maybeN}}
$

\vsepRule


$\inferrule*[lab=\ruleNameFig{T-Fun}]
  {
   \varUnTyp=\utArrowThree
                {\seq{\tyVar}}{\seq{\varLoc}}{\seq{\varHeap}}
                {\triWorld{x}{\varTyp_1}{\heapSig_1}}
                {\varWorld_2} \sepPremise
   \relWf{\Gamma}{\varUnTyp} \sepPremise
   \labelEnv_1 = \empLabelEnv \\\\
   \applyHelperOp{HeapEnv}{\heapSig_1} =
     (\tyBind{\seq{z}}{\seq{\varScm}}, \heapTyp_1) \sepPremise
   \Gamma_1 = \Gamma,
              \seq{\tyVar}, \seq{\varLoc}, \seq{\varHeap},
              \tyBind{x}{\varTyp_1}, \tyBind{\seq{z}}{\seq{\varScm}} \\\\
   \relTypExpLabel
             {\Gamma_1}{\heapTyp_1}{\labelEnv_1}
             {e}{\varTyp_2}{\heapTyp_2}{\maybeN} \sepPremise
   \relWorldSat{\Gamma_1}{\world{\varTyp_2}{\heapTyp_2}}
               {\varWorld_2}{\varSubst}{\maybeN}
  }
  {\relTypVal{\Gamma}{\heapTyp}
             {\funBare{x}{e}}
             {\refTypX{y}{y=\funBare{x}{e}\wedge\hasTyp{y}{\varUnTyp}}}
             {\maybeN}}
$

\vsepRule


\judgementHead{World Satisfaction}
              {$\relWorldSat{\Gamma}{\world{\varTyp}{\heapTyp}}
                            {\varWorld}{\varSubst}{\maybeN}$}

\vsepRule



$\inferrule*
  {
   \relSub{\Gamma}{\varTyp}{\varScm}{\maybeN} \sepPremise
   \relMatch{\heapTyp}{\heapSig}{\varSubst} \sepPremise
   \relImplForm{\Gamma, \tyBind{x}{\varTyp}}{\varSubst\embed{\heapSig}}
  }
  {\relWorldSat{\Gamma}
               {\world{\varTyp}{\heapTyp}}
               {\world{\tyBind{x}{\varScm}}{\heapSig}}
               {\varSubst'}{\maybeN}}
$

\vsepRule

\input{fig-dec-typing-values-more}

\caption{Value type checking for \dImp}
\label{fig:dec-typing-values}

\end{figure}

We supplement our discussion of value typing, defined in
\autoref{fig:dec-typing-values}, from \refSectypechecking.
The $\ruleName{T-Extend}$ rule for dictionaries is straightforward.
The $\ruleName{T-Loc}$ rule assigns run-time location $\runLoc$ (which appears
during evaluation, but not in source programs) a reference type corresponding
to its compile-time location, using the mapping $\helperOp{StaticLoc}$.
Notice that, unlike the version of the rule from
\refFigdectypingvaluesselected,
the rule $\ruleName{T-Fun}$ uses an empty label environment to type
check function bodies, so that break expressions cannot cross
function boundaries.

\subsection{Expression Typing}

\begin{figure*}[t]
\figureFontSize
\centering

\judgementHead{Expression Typing}
  {$\relTypExp{\Gamma}{\heapTyp}{e}{\varTyp}{\heapTyp'}{\maybeN}$}

\vsepRule

\textrm{Rules from \refFigdectypingexpselected, updated with
  label environments:}\hfill{ }

\vsepRule

$\inferrule*[left=\ruleNameFig{T-Val}]
  {\relTypVal{\Gamma}{\heapTyp}{\varVal}{\varTyp}{\maybeN}}
  {\relTypExp{\Gamma}{\heapTyp}{\varVal}{\varTyp}{\heapTyp}{\maybeN}}
$
\hsepRule
$\inferrule*[right=\ruleNameFig{T-Let}]
  {\relTypExp{\Gamma}{\heapTyp}
             {e_1}{\varTyp_1}{\heapTyp_1}{\maybeN} \sepPremise
   \relTypExp{\Gamma,\tyBind{x}{\varTyp_1}}{\heapTyp_1}
             {e_2}{\varTyp_2}{\heapTyp_2}{\maybeN}
  }
  {\relTypExp{\Gamma}{\heapTyp}
             {\letinBare{x}{e_1}{e_2}}
             {\existsTyp{x}{\varTyp_1}{\varTyp_2}}
             {\heapTyp_2}{\maybeN}}
$

\vsepRule

$\inferrule*[right=\ruleNameFig{T-If}]
  {\relTypVal{\Gamma}{\heapTyp}{\varVal}{\varScm}{\maybeN} \sepPremise
   \relTypExp{\Gamma,\truthy{\varVal}}{\heapTyp}
             {e_1}{\varTyp_1}{\heapTyp_1}{\maybeN} \sepPremise
   \relTypExp{\Gamma,\falsy{\varVal}}{\heapTyp}
             {e_2}{\varTyp_2}{\heapTyp_2}{\maybeN} \sepPremise
   \world{\varTyp}{\heapTyp'} = \iteJoin{\varVal}
                                     {\world{\varTyp_1}{\heapTyp_1}}
                                     {\world{\varTyp_2}{\heapTyp_2}}
  }
  {\relTypExp{\Gamma}{\heapTyp}
             {\ite{\varVal}{e_1}{e_2}}
             {\varTyp}{\heapTyp'}{\maybeN}}
$

\vsepRule

$\inferrule*[left=\ruleNameFig{T-Ref}]
  {\locName\notin\domR{\heapTyp} \sepPremise
   \relTypVal{\Gamma}{\heapTyp}{\varVal}{\varTyp}{\maybeN} \sepPremise
   \heapTyp' = \heapCat{\heapTyp}{\heapBind{\locName}{\varVal}}
  }
  {\relTypExp{\Gamma}{\heapTyp}
             {\newref{\locName}{\varVal}}
             {\utRef{\locName}}{\heapTyp'}{\maybeN}}
$
\hsepRule
%
$\inferrule*[right=\ruleNameFig{T-Deref}]
  {\relTypVal{\Gamma}{\heapTyp}{\varVal}
             {\typRef{\locName}}
             {\maybeN} \sepPremise
   \heapEq{\heapTyp}
          {\heapCat{\heapTyp_0}
                   {\heapBind{\locName}{\varVal'}}}
  }
  {\relTypExp{\Gamma}{\heapTyp}{\ \deref{\varVal}}
             {\refTypX{y}{y=\varVal'}}{\heapTyp}{\maybeN}}
$

\vsepRule

$\inferrule*[left=\ruleNameFig{T-Setref}]
  {
   \relTypVal{\Gamma}{\heapTyp}{\expTupleTwo{\varVal_1}{\varVal_2}}
             {\tyTupleTwo{\utRef{\locName}}{\varTyp}}{\maybeN} \\\\
   \heapEq{\heapTyp}
          {\heapCat{\heapTyp_0}{\heapBind{\locName}{\varVal}}} \sepPremise
   \heapTyp'=
     \heapCat{\heapTyp_0}
             {\heapBind{\locName}{\varVal_2}}
  }
  {\relTypExp{\Gamma}{\heapTyp}{\setref{\varVal_1}{\varVal_2}}
             {\refTypX{x}{x=\varVal_2}}{\heapTyp'}{\maybeN}}
$
\hsepRule
$\inferrule*[right=\ruleNameFig{T-NewObj}]
  {\locName_1 \notin \domR{\heapTyp} \sepPremise
   \relTypVal{\Gamma}{\heapTyp}{\expTupleTwo{\varVal_1}{\varVal_2}}
             {\tyTupleTwo{\tyDict}{\utRef{\locName_2}}}{\maybeN} \\\\
   \heapEq{\heapTyp}
          {\heapCat{\heapTyp_0}
                   {\heapBindPair{\locName_2}{\varVal'}{\locName_3}}} \sepPremise
   \heapTyp' =
      \heapCat{\heapTyp}
              {\heapBindPair{\locName_1}{\varVal_1}{\locName_2}}
  }
  {\relTypExp{\Gamma}{\heapTyp}{\newObjThree{\locName_1}{\varVal_1}{\varVal_2}}
             {\utRef{\locName_1}}{\heapTyp'}{\maybeN}}
$

\vsepRule

$\inferrule*[right=\ruleNameFig{T-App}]
  {
   \relTypVal
      {\Gamma}{\heapTyp}{\varVal_1}
         {\utArrowThree
             {\seq{\tyVar}}{\seq{\varLoc}}{\seq{\varHeap}}
             {\varWorld_{1}}{\varWorld_{2}}}{\maybeN} \sepPremise
   \relTypVal{\Gamma}{\heapTyp}{\varVal_2}{\varTyp_2}{\maybeN} \sepPremise
   \relWf{\Gamma}{\substOne{\seq{\tyVar}}{\seq{\varTyp}}} \sepPremise
   \relWf{\Gamma}{\substOne{\seq{\varLocEither}}{\seq{\eitherLocName}}} \sepPremise
   \relWf{\Gamma}{\substOne{\seq{\varHeap}}{\seq{\heapSig}}} \\\\
   \varWorld_{2}' = \freshen{\varWorld_{2}} \sepPremise
   (\varWorld_{1}', \varWorld_{2}'') =
     \expand
       {\hInst{\lInst{\tInst{(\varWorld_{1}, \varWorld_{2}')}
          {\seq{\tyVar}}{\seq{\varTyp}}}
          {\seq{\varLoc}}{\seq{\locName}}}
          {\seq{\varHeap}}{\seq{\heapSig}}} \\\\
   \relWorldSat{\Gamma}{\world{\varTyp_2}{\heapTyp}}
               {\varWorld_{1}'}{\varSubst}{\maybeN} \sepPremise
   \varWorld_{1}' = \triWorld{x}{\varTyp_{11}}{\heapSig_{11}} \sepPremise
   \varSubst' = \varSubst\substOne{x}{\varVal_2} \sepPremise
   \varSubst'\varWorld_{2}'' =
      \triWorld{x'}{\varTyp_{12}}{\heapSig_{12}} \sepPremise
   \applyHelperOp{HeapEnv}{\heapSig_{12}} =
     (\tyBind{\seq{y}}{\seq{\varScm}}, \heapTyp_{12})
  }
  {\relTypExp{\Gamma}{\heapTyp}
             {\appThree{\seq{\varTyp}}{\seq{\locName}}{\seq{\heapSig}}
                {\varVal_1}{\varVal_2}}
             {\existsTyp{x'}{\varTyp_{12}}{
              \existsTyp{\seq{y}}{\seq{\varScm}}{\refTypX{z}{z=x'}}}}
             {\heapTyp_{12}}{\maybeN}}
$

\vsepRule

\textrm{Additional rules:}\hfill{ }

\vsepRule

$\inferrule*[left=\ruleNameFig{T-As}]
  {\relWf{\Gamma}{\varTyp} \sepPremise
   \relTypExp{\Gamma}{\heapTyp}{e}{\varTyp}{\heapTyp}{\maybeN}}
  {\relTypExp{\Gamma}{\heapTyp}{\cast{e}{\varTyp}}{\varTyp}{\heapTyp}{\maybeN}}
$
\hsepRule
$\inferrule*[right=\ruleNameFig{T-Sub}]
  {
   \relTypExp{\Gamma}{\heapTyp}{e}{\varScm}{\heapTyp'}{\maybeN} \sepPremise
   \relSub{\Gamma}{\varScm}{\varTyp}{\maybeN} \sepPremise
   \relWf{\Gamma}{\varTyp}
  }
  {\relTypExp{\Gamma}{\heapTyp}{e}{\varTyp}{\heapTyp'}{\maybeN}}
$

\vsepRule

$\inferrule*[lab=\ruleNameFig{T-Freeze}]
  {\relTypVal{\Gamma}{\heapTyp}{\varVal}
                     {\typRef{\locName}}{\maybeN} \sepPremise
   \Gamma(\weakLocName) = (\varTyp, \locName') \\\\
   \heapEq{\heapTyp}
          {\heapCatThr{\heapTyp_0}
                      {\heapBind{\weakLocName}{\thawState}}
                      {\heapBindPair{\locName}{\varVal'}{\locName'}}} \sepPremise
   \thawState=\frzn \textrm{ or } \thawState=\thwd{\locName} \\\\
   \relTypVal{\Gamma}{\heapTyp}{\varVal'}{\varTyp}{\maybeN} \sepPremise
   \heapTyp' = \heapCat{\heapTyp_0}{\heapBind{\weakLocName}{\frzn}}
  }
  {\relTypExp{\Gamma}{\heapTyp}
             {\freeze{\weakLocName}{\varVal}}
             {\refTyp{\hasTyp{\theV}{\utRef{\weakLocName}}\wedge
                      \theV\neq\vNull}}
             {\heapTyp'}
             {\maybeN}}
$
%
%
%
\hsepRule
$\inferrule*[lab=\ruleNameFig{T-Thaw}]
  {
   \relTypVal{\Gamma}{\heapTyp}{\varVal}{\typRef{\weakLocName}}{\maybeN} \sepPremise
   \Gamma(\weakLocName) = (\varTyp, \locName') \\\\
   \heapEq{\heapTyp}{\heapCat{\heapTyp_0}{\heapBind{\weakLocName}{\frzn}}} \sepPremise
   \heapTyp'=
      \heapCatThr{\heapTyp_0}
                 {\heapBind{\weakLocName}{\thwd{\locName}}}
                 {\heapBindPair{\locName}{x}{\locName'}} \\\\
   \varScm = \refTypX{y}{\formIte{(v=\vNull)}{(y=\vNull)}
                                 {(\hasTyp{y}{\utRef{\locName}})}}
  }
  {\relTypExp{\Gamma}{\heapTyp}
             {\thaw{\locName}{\varVal}}
             {\existsTyp{x}{\varTyp}{\varScm}}
             {\heapTyp'}
             {\maybeN}}
$

\vsepRule

$\inferrule*[left=\ruleNameFig{T-Label}]
  {\labelEnv' = 
     \labelEnv, \tyBind{\labelNameX{\labelName}}{(\world{\varTyp}{\heapSig})} \sepPremise
   \relTypExpLabel
      {\Gamma}{\heapTyp}{\labelEnv'}{e}{\varTyp}{\heapTyp'}{\maybeN} \\\\
   \relHeapSat{\Gamma}{\heapTyp'}{\heapSig}{\varSubst}{} \sepPremise
   \applyHelperOp{HeapEnv}{\heapSig} =
     (\tyBind{\seq{x}}{\seq{\varScm}}, \heapTyp'')
  }
  {\relTypExp{\Gamma}{\heapTyp}{\expLabel{\labelName}{e}}
             {\existsTyp{\seq{x}}{\seq{\varScm}}{\varTyp}}{\heapTyp''}{\maybeN}}
$
\hsepRule
$\inferrule*[right=\ruleNameFig{T-Break}]
  {\labelEnv(\labelNameX{\labelName}) = \world{\varTyp}{\heapSig} \\\\
   \relTypVal{\Gamma}{\heapTyp}{\varVal}{\varTyp}{} \sepPremise
   \relHeapSat{\Gamma}{\heapTyp}{\heapSig}{\varSubst}{}
  }
  {\relTypExp{\Gamma}{\heapTyp}{\expBreak{\labelName}{\varVal}}
             {\refTyp{\formFalse}}{\heapTyp}{\maybeN}}
$

\vsepRule

\caption{Expression type checking for \dImp}
\label{fig:dec-typing-all-exp}

\end{figure*}

When we presented expression typing in \refSectypecheckingexp, we
ignored break and label expressions, so the typing judgement referred only
to type and heap environments. To account for control operators, the expression
typing judgement is of the form
$\relTypExp{\Gamma}{\heapTyp}{e}{\varTyp}{\heapTyp'}{}$,
where a label environment is an additional input.
We define the typing rules in \autoref{fig:dec-typing-all-exp} and
supplement our previous discussion.
The $\ruleName{T-As}$ and $\ruleName{T-Sub}$ rules are straightforward.
Aside from the rules for label and break expressions, label environments
$\labelEnv$ play no interesting role.
The rules we discussed in \refSectypecheckingexp\ carry over directly
to the formulation with label environments.
%

\parahead{Weak Location Bindings}

For simplicity, we assume that the initial type environment contains all the
weak location bindings $\heapBindPair{\weakLocName}{\varTyp}{\locName}$
required by the program.

\parahead{Thaw and Freeze}

To safely allow a weak location $\weakLocName$ to be treated
\emph{temporarily} as strong, $\dRef$ ensures that $\weakLocName$ has at
most one corresponding \emph{thawed} location at a time; if there is none,
we say $\weakLocName$ is \emph{frozen}.
The rule $\ruleName{T-Thaw}$ thaws $\weakLocName$ to a strong location
$\locName$ (which we syntactically require be distinct from all other
thawed locations for $\weakLocName$) and updates the heap environment
with thaw state $\thwd{\locName}$ to track the correspondence.
Subtyping allows $\vNull$ weak references, so the output type
is $\vNull$ if the original reference is; otherwise, it is a reference
to $\locName$. Finally, the new heap also binds a value $x$ of type
$\varTyp$, the invariant for all values stored at $\weakLocName$, and
the output type introduces an \emph{existential} so that $x$ is in scope
in the new heap.

The rule $\ruleName{T-Freeze}$ serves two purposes, to merge a
strong location $\locName$ into a weak (frozen) location $\weakLocName$
and to \emph{re-freeze} a thawed (strong) location $\locName$ that
originated from $\weakLocName$,
as long as the heap value stored at $\locName$ satisfies the invariant
required by $\weakLocName$.
The strong reference is guaranteed to be non-$\vNull$, so the output
type remembers that the frozen reference is, too.
Compared to the presentation in~\cite{LinLocFI07}, we have combined
freeze and re-freeze into a single \verb+freeze+ expression
that includes an explicit thaw state $\thawState$.

The result of thawing a weak location is \emph{either} a strong reference
or $\vNull$.
Although we could statically require that all strong references be
non-$\vNull$ before use (to rule out the possibility of null-dereference
exceptions), we choose to allow $\vNull$ references to facilitate
idiomatic programming.
Therefore, we modify the input type for the object primitives in
\verb+objects.dref+ to allow a $\vNull$ argument.
For example, consider the updated input type for \verb+hasPropObj+ below, where
$\tyMaybeNull{\varTyp} \defeq \refTyp{\varTyp(\theV)\vee\theV=\vNull}$.
Notice that we add the predicate $x\neq\vNull$ to the \emph{output} type,
because if \verb+hasPropObj+ evaluates without raising an exception, then $x$
is guaranteed to be non-$\vNull$.
In this way, $\dRef$ precisely tracks the invariants of thawed
objects (\cf \verb+passengers+ from \refSecoverviewcollections).

\equationSizeBegin
\begin{align*}
&\
  \world
    {\tyTupleTwo{\tyBind{x}{\tyMaybeNull{\refCon}}}{\tyBind{k}{\tyStr}}}
    {\heapBindPair{x}{\tyBind{d}{\tyDict}}{\locProto{x}}} \\[-2pt]
\ttArr &\
  \world
    {\refTypShort{x\neq\vNull\wedge
                  (\formIff{\theV}{\objHasR{d}{k}{\curHeap}{\locProto{x}}})}}
    {\sameHeap}
\end{align*}
\equationSizeEnd

\parahead{Type Instantiation}

The $\helperOp{TInst}$ procedure processes
has-type predicates in formulas as follows:
\equationSizeBegin
\begin{align*}
 \instantiate{\hasTyp{\varLogVal}{\tyvar}}{\tyvar}{\refTypX{x}{\varFormOne}}
   &= \subst{\varFormOne}{x}{\varLogVal}
 \\
 \instantiate{\hasTyp{\varLogVal}{\tyvarB}}{\tyvar}{\varTyp}
   &= \hasTyp{\varLogVal}{\tyVarB}
\end{align*}
\equationSizeEnd

\parahead{Join}

The \ruleName{T-If} rule uses a \helperOp{Join} operator, defined in
\autoref{fig:join}, that combines the type and heap environments along each
branch ($\world{\varTyp_1}{\heapTyp_1}$ and $\world{\varTyp_2}{\heapTyp_2}$)
such that the type and output heap for the overall if-expression
($\world{\varTyp}{\heapTyp'}$) are in prenex form.
The operator starts by using \helperOp{JoinTypes} to move existential
binders for the types to the top-level. Rearranging variables in this way is
sound because we assume that, by convention, all let-bound variables in a
program are distinct.
Then, we use \helperOp{JoinHeaps} to combine the bindings in a heap
environment one location at a time.
We show a few representative equations in \autoref{fig:join},
abusing notation in several ways. For example, we write
$\varHeapBinding\ \backslash\ \locName$ to denote that $\locName$ is
not bound in $\varHeapBinding$.
When a location $\locName$ is bound in both heaps to values
$\varVal_1$ and $\varVal_2$, respectively, \helperOp{JoinHeaps} introduces
a new binding $y$ whose type is the join of $\varVal_1$ and $\varVal_2$.
When a location $\locName$ is bound in only one heap, we use the dummy type
$\tyTop$ to describe the (non-existent) value in the other heap.
There is no danger that $\locName$ will be unsoundly dereferenced after
the if-expression, since \helperOp{JoinTypes} guards the types of references
$\utRef{\locName}$ with the appropriate guard predicates.

\begin{figure*}
\equationSizeBegin
\begin{align*}
\iteJoin{b}{\world{\varScm_1}{\heapTyp_1}}{\world{\varScm_2}{\heapTyp_2}} = &\
  \world
    {\existsTyp{\seq{x}}{\seq{\varTyp}}
               {\existsTyp{\seq{y}}{\seq{\varTyp'}}
                          {\varScm}}}
    {\heapTyp}
  \hspace{0.10in}\textrm{where }
  \iteJoinTypes{b}{\varScm_1}{\varScm_2} =
    \existsTyp{\seq{x}}{\seq{\varTyp}}{\varScm}
\\[-4pt] &\
\phantom{
  \world
    {\existsTyp{\seq{x}}{\seq{\varTyp}}
               {\existsTyp{\seq{y}}{\seq{\varTyp'}}
                          {\varScm}}}
    {\heapTyp}
}
  \hspace{0.10in}\textrm{and }
  \iteJoinHeaps{b}{\heapTyp_1}{\heapTyp_2} =
    (\exists \seq{y}:\seq{\varTyp'}, \heapTyp)
\\[4pt]
\iteJoinTypes{b}{\varScm_1}{\varScm_2} = &\
  \refTyp{(b=\vTrue\Rightarrow\embed{\varScm_1}) \wedge
          (b=\vFalse\Rightarrow\embed{\varScm_2})}
\\
\iteJoinTypes{b}{(\existsTyp{\seq{x_1}}{\seq{\varTyp_1}}{\varScm_1})}{\varScm_2} = &\
  \existsTyp{\seq{x_1}}
            {(\iteJoinTypes{b}{\seq{\varTyp_1}}{\seq{\tyTop}})}
            {\iteJoinTypes{b}{\varScm_1}{\varScm_2}}
\\
\iteJoinTypes{b}{\varScm_1}{(\existsTyp{\seq{x_2}}{\seq{\varTyp_2}}{\varScm_2})} = &\
  \existsTyp{\seq{x_2}}
            {(\iteJoinTypes{b}{\seq{\tyTop}}{\seq{\varTyp_2}})}
            {\iteJoinTypes{b}{\varScm_1}{\varScm_2}}
\\
\iteJoinTypes{b}{(\existsTyp{\seq{x_1}}{\seq{\varTyp_1}}{\varScm_1})}
                 {(\existsTyp{\seq{x_2}}{\seq{\varTyp_2}}{\varScm_2})} = &\
  \existsTyp{\seq{x_1}}{(\iteJoinTypes{b}{\seq{\varTyp_1}}{\seq{\tyTop}})}{
  \existsTyp{\seq{x_2}}{(\iteJoinTypes{b}{\seq{\tyTop}}{\seq{\varTyp_2}})}{
     \iteJoinTypes{b}{\varScm_1}{\varScm_2}}}
\\[6pt]
\iteJoinHeaps{b}{\heapCat{\heapBind{\locName}{\varVal_1}}{\varHeapBinding_1}}
                {\heapCat{\heapBind{\locName}{\varVal_2}}{\varHeapBinding_2}} = &\
  (\exists \tyBind{y}{\iteJoinTypes{b}{\refTypX{x}{x=\varVal_1}}
                                      {\refTypX{x}{x=\varVal_2}}},
   \heapCat{\heapBind{\locName}{y}}
           {\iteJoinHeaps{b}{\varHeapBinding_1}{\varHeapBinding_2}})
\\
\iteJoinHeaps{b}{\heapCat{\heapBind{\locName}{\varVal_1}}{\varHeapBinding_1}}
                {\varHeapBinding_2 \backslash\ \locName} = &\
  (\exists \tyBind{y}{\iteJoinTypes{b}{\refTypX{x}{x=\varVal_1}}{\tyTop}},
   \heapCat{\heapBind{\locName}{y}}
           {\iteJoinHeaps{b}{\varHeapBinding_1}{\varHeapBinding_2}})
\\
\iteJoinHeaps{b}{\varHeapBinding_1 \backslash\ \locName}
                {\heapCat{\heapBind{\locName}{\varVal_2}}{\varHeapBinding_2}} = &\
  (\exists \tyBind{y}{\iteJoinTypes{b}{\tyTop}{\refTypX{x}{x=\varVal_2}}},
   \heapCat{\heapBind{\locName}{y}}
           {\iteJoinHeaps{b}{\varHeapBinding_1}{\varHeapBinding_2}})
\end{align*}
\caption{Environment join}
\label{fig:join}
\end{figure*}

\parahead{Control Operators}

The \ruleName{T-Label} rule for $\expLabel{x}{e}$ binds the label
$\labelNameX{x}$ to an expected world $\world{\varTyp}{\heapSig}$ in the
label environment $\labelEnv'$ used to check $e$, and expects
that \emph{all} exit points of $e$ produce a value and heap environment
that satisfy the expected world.
The exit points are all  $\expBreak{\labelName}{\varVal}$ expressions in $e$,
as well as the ``fall-through" of expression $e$ for control flow paths that
do not end with \verb+break+;
the \ruleName{T-Break} rule handles the former cases, and
the second and third premises of \ruleName{T-Label} handle the latter.
If all exit points satisfy the expected world, we use the
\helperOp{HeapEnv} procedure to convert the heap type into a
heap environment, like in the \ruleName{T-App} rule.
Notice that \ruleName{T-Break} derives the type $\refTyp{\formFalse}$
because a \verb+break+ immediately completes the evaluation context,
thus making the subsequent program point unreachable.

\subsection{Desugaring}

\begin{figure}[t]
\figureFontSize

\judgementHeadThree{Desugaring}{(extends \refFigdesugaring)}
              {$\dsLHS{e}\desugarSym e$}

\vsepRule

$\dsLHS{function\ (\seq{x})\ \dsAnn{\mathit{\varTyp}}\ \{\ e\ \}}$
     \desugarSym \
\dsRuleNameFig{DS-Func}

\gap

  \hsepRule
  ${\lambda (\mathit{this},\mathit{arguments}).}$

    \hsepRule \sep\sep
    ${\letinBare
        {\expTupleTwo{\dsVar{\mathtt{x_0}}}{\ldots}}
        {\expTupleTwo
           {\newref{\locConst_{x_0}}{(\vGet\ \mathit{arguments}\ \ttfld{0})}}
           {\ldots}}
        {}}$

    \hsepRule \sep\sep
    ${\expLabel{\mathit{return}}{\ds{e}}}$

\gap

$\dsLHS{function\ F(\seq{x})\ \dsAnn{\mathit{\mathtt{\#}\newAnn\ \varTyp}}\ \{\ e\ \}}$
     \desugarSym \
\dsRuleNameFig{DS-Ctor}

\gap

  \hsepRule
  $\texttt{let }{\mathit{f}}={\lambda (\mathit{this},\mathit{arguments}).\ }$

    \hsepRule \sep\sep
    ${\letinBare
        {\expTupleTwo{\dsVar{\mathtt{x_0}}}{\ldots}}
        {\expTupleTwo
           {\newref{\locConst_{x_0}}{(\vGet\ \mathit{arguments}\ \ttfld{0})}}
           {\ldots}}
        {}}$

    \hsepRule \sep\sep
    ${\expLabel{\mathit{return}}{\ds{e}}}\ \mathtt{ in}$

  \hsepRule
  $\letinBare
     {\mathit{p}}
     {\newObjThree{\locConst_{\mathit{Fproto}}}{\emptydict}{(\dsGetProto{\mathtt{Object}})}}
     {}$

  \hsepRule
  $\letinBare
     {\mathit{d}}
     {\{\ \ttfld{\_\_code\_\_}=\mathit{f}\ \texttt{as}\ \ds{\mathit{\varTyp}};\
          \ttfld{prototype}=\mathit{p}\ \}}
     {}$

  \hsepRule
  $\newObjThree{\locConst_{F}}{\mathit{d}}{(\dsGetProto{\mathtt{Function}})}$

\gap

$\dsLHS{return\ e} \desugarSym\ \expBreak{\mathit{return}}{\ds{e}}$
\dsRuleNameFig{DS-Return}

\gap

$\dsLHS{\dsAnn{\mathit{\varTyp}}\ 
         while\ (e_{cond})\ \{\ e_{body}\ \}}$ \desugarSym\
\dsRuleNameFig{DS-While}

\gap

  \hsepRule
  $\labelNameX{break} :
     \texttt{letrec}\ \hasTyp{\mathit{loop}}{\varTyp} = \lambda ().$

    \hsepRule \hsepRule \hsepRule \hsepRule
    $\texttt{if } {\ds{e_{cond}}}
     \texttt{ then } {({\ds{e_{body}}}; \mathit{loop\ ()})}$

    \hsepRule \hsepRule \hsepRule \hsepRule
    $\texttt{else } \vUndefined\ \texttt{in}\ \mathit{loop\ ()}$

\gap

\dsLHS{break} \desugarSym\
$\expBreak{\mathit{break}}{\vUndefined}$
\dsRuleNameFig{DS-Break}

\gap

\dsLHS{\dsAnn{\#thaw\ \locName\ e}} \desugarSym \
\thaw{\locName}{\ds{e}}
\dsRuleNameFig{DS-Thaw}

\gap

\dsLHS{\dsAnn{\#freeze\ \weakLocName\ \thawState\ e}} \desugarSym \
\freeze{\weakLocName}{\ds{e}}
\dsRuleNameFig{DS-Freeze}

\gap

\dsLHS{assert(e)} \desugarSym \ \ds{e}\ \texttt{as}\ \refTyp{\theV=\vTrue}
\dsRuleNameFig{DS-Assert}

\gap

\caption{Desugaring $\djs$ to $\dRef$}
\label{fig:desugaring-more}
\end{figure}

In \autoref{fig:desugaring-more}, we show more of the desugaring rules.

\parahead{Functions and Constructors}

As discussed in \refSecdesugaring, we desugar non-constructor
functions (\ruleName{DS-Func}) to scalar function values and
constructor functions (\ruleName{DS-Ctor}) to objects.
Following $\LamJS$~\cite{Guha10a}, we wrap each desugared function body with the
label $\labelNameX{\mathit{return}}$, which facilitates the
desugaring of \verb+return+ statements (\ruleName{DS-Return}).
We desugar named, recursive $\djs$ functions via the standard \verb+letrec+
encoding using \verb+fix+; we omit this rule from \autoref{fig:desugaring-more}.

\ruleName{DS-Ctor} first creates a fresh object at location $\locConst_{F}$
with prototype \verb+Function.prototype+,
then stores the desugared constructor function in the \ttfld{\_\_code\_\_}
field, and finally creates an empty object at location $\locConst_{F_{proto}}$
that is stored in the \ttfld{prototype} field, to be used when creating an
object with this constructor (\ruleName{DS-New}).

\parahead{Loops}

Following \LamJS, the \ruleName{DS-While} desugars \verb+while+ loops
to recursive functions (we write \verb+letrec+ as syntactic sugar for the
standard encoding using \verb+fix+).
As such, a (function type) annotation describes the invariants
that hold before and after each iteration.
A $\labelNameX{\mathit{break}}$ label around the desugared loop body
facilitates the desugaring of \verb+break+ statements (\ruleName{DS-Break}).
We elide similar mechanisms for \verb+do-while+ loops, \verb+for+ loops,
\verb+for-in+ loops, and \verb+continue+ statements.


\section{Extensions}
\label{sec:appendix-extensions}

We now outline two ways to increase the expressiveness of location
polymorphism in $\dRef$.


\subsection{Weak Location Polymorphism}

So far, we have universally quantified function types over strong locations.
We can make several changes to allow quantification over weak locations as well.
First, we extend the syntax of locations.

\equationSizeBegin
$$
{\weakLocName}\sep\sep{::=}\sep\sep{\cdots}\spaceItemSmall{\locFrozen{\varLoc}}
\sep\sep\sep\sep\sep
{\eitherLocName}\sep\sep{::=}\sep\sep{\locName}\spaceItemSmall{\weakLocName}
\sep\sep\sep\sep\sep
{\varLocEither}\sep\sep{::=}\sep\sep{\varLoc}\spaceItemSmall{\locFrozen{\varLoc}}
$$
\equationSizeEnd
We use $\locFrozen{\varLoc}$ to range over weak location variables, and
we extend the grammar of weak locations $\weakLocName$ to include them
(in addition to weak location constants $\locFrozen{\locConst}$).
We also define $\eitherLocName$ (resp. $\varLocEither$) to range over
\emph{arbitrary} locations (resp. location variables).
Next, we extend the syntax of function types and function application.

\equationSizeBegin
$$
\begin{array}{rrcll}
\startRow{\textbf{}}{e}{::=}
  {\cdots}
  \spaceItem
  {\appThree{\seq{\varTyp}}{\seq{\eitherLocName}}{\seq{\heapSig}}{\varVal_1}{\varVal_2}}
\spaceCategory
\startRow{\textbf{}}{\varUnTyp}{::=}
  {\cdots}
  \spaceItem
  {\utArrowDrefWorlds
     {\seq{\tyVar}}{\seq{\varLocEither}}{\seq{\varHeap}}
     {\world{\weakHeap}{\varWorld_1}}{\varWorld_2}}
\spaceCategory
\startRow{\textbf{}}{\weakHeap}{::=}
  {\heapBindPair{\weakLocName}{\varTyp}{\locName}}
  \spaceItem
  {\heapCat{\weakHeap_1}{\weakHeap_2}}
  \spaceItem
  {\emptyset}
\end{array}
$$
\equationSizeEnd
A function type is now parametrized over arbitrary location variables
$\seq{\varLocEither}$ and a \emph{weak heap} $\weakHeap$ of bindings
that describe weak location variables.
To match, function application now includes location arguments
$\seq{\eitherLocName}$ rather than $\seq{\locName}$;
typing must ensure that strong location variables
$\varLoc$ (resp. weak location variables $\locFrozen{\varLoc}$) are instantiated
only with strong locations $\locName$ (resp. weak locations $\weakLocName$).

A function type refers to a weak heap only in the domain of the function,
because weak locations are flow-insensitive and do not vary at different
program points.
Before, we assumed that the initial typing environment contained bindings
for all weak locations. The new syntax of function types replaces this
convention by \emph{abstracting} over weak locations.
Consequently, the function application rule must check that the declared
weak heap $\weakHeap$ of a function type is satisfiable given the current
heap environment $\heapTyp$ at a call site
(after substitution of all polymorphic variables).


\subsection{Existential Locations}

Universally quantifying over all locations, including simple locations,
clutters function types and applications with additional arguments, and also
exposes locations that are ``internal" or ``local" to the desugared $\dRef$
program and \emph{not} accessible in the original $\djs$ program.

Consider the following example; we refer to the original function as $f$
and the desugared version as $f'$.

\begin{figure}[h]
\begin{minipage}[b]{0.30\linewidth}
\centering
\input{exists-loc.js}
\end{minipage}
\hspace{0.0cm}
\begin{minipage}[b]{0.66\linewidth}
\centering
\input{exists-loc.dref}
\end{minipage}
\end{figure}

\noindent
We might annotate the $\djs$ function $f$ with the type

\equationSizeBegin
\begin{align*}
& \forall \varLoc,\varLoc'.\
\world{\utRef{\varLoc}}
      {\heapBindPair{\varLoc}{\tyDict}{\varLoc'}}
\\[-2pt]
\ttArr\ &
\world{\tyTop}
      {\heapBind{\varLoc}{\sameHeap}}
\intertext{\normalsize and the desugared version would have the type}
& \forall \varLoc,\varLoc',\varLoc_x,\varLoc_y.\
\world{\utRef{\varLoc}}
      {\heapBindPair{\varLoc}{\tyDict}{\varLoc'}}
\\[-2pt]
\ttArr\ &
\world{\tyTop}
      {\heapCatThr{\heapBind{\varLoc}{\sameHeap}}
                  {\heapBind{\varLoc_x}{\utRef{\varLoc}}}
                  {\heapBind{\varLoc_y}{\utRef{\varLoc}}}}
\end{align*}
\equationSizeEnd

\noindent
that uses additional location variables for the references inserted by
the translation.
Although it is straightforward to mechanically desugar function types in
this manner, the additional location parameters at function calls increase
the manual annotation burden or, more likely since we cannot expect $\djs$
programmer to write them, the burden on the type system to infer them.

Instead, we can introduce \emph{existential location types} into the system
and write the following type for the desugared function $f'$.

\equationSizeBegin
\begin{align*}
& \forall \varLoc,\varLoc'.\
\world{\utRef{\varLoc}}
      {\heapBindPair{\varLoc}{\tyDict}{\varLoc'}}
\\[-2pt]
\ttArr\ &
\world{\existsLoc{\varLoc_x,\varLoc_y}{\tyTop}}
      {\heapCatThr{\heapBind{\varLoc}{\sameHeap}}
                  {\heapBind{\varLoc_x}{\utRef{\varLoc}}}
                  {\heapBind{\varLoc_y}{\utRef{\varLoc}}}}
\end{align*}
\equationSizeEnd

\noindent
Notice that the \emph{output} world uses existentials to name the
(strong, simple) locations inserted by desugaring. As a result,
a call to this function need not instantiate the local locations;
instead, the type system can generate \emph{fresh} location
constants (\ie skolemize) for the existential locations.

We provide a sketch of how to extend $\dRef$ with existential locations.
First, we extend the syntax of types.

\equationSizeBegin
$$
\begin{array}{rrcll}
\startRow{\textbf{}}{\varTyp}{::=}
  {\cdots} \spaceItem {\existsLoc{\locName}{\varTyp}}
\end{array}
$$
\equationSizeEnd
We intend that existential locations only appear in \emph{positive}
positions of function types, which we can compute in similar fashion
to the \helperOp{Poles} procedure from $\dtypes$~\cite{NestedPOPL12}
that tracks \emph{polarity} of types nested within formulas.
Effectively, we require that every function type be of the form

\equationSizeBegin
$$\utArrowDrefWorlds{\seq{\tyVar}}{\seq{\varLoc}}{\seq{\varHeap}}
   {\triWorld{x}{(\refTypX{y}{\varFormOne})}{\heapSig}}
   {\triWorld{x'}{(\existsLoc{\seq{\locName}}{\refTypX{y'}{\varFormOne'}})}
             {\heapSig'}}$$
\equationSizeEnd

\noindent
where the input type is a refinement type and the output type is in a
\emph{prenex} form
that requires all existentially-quantified locations to appear at the
top-level and which prohibits existentially-quantified values.
Intuitively, the locations $\seq{\locName}$ correspond to local reference
cells that a function allocates when invoked and are inaccessible to callers.

To introduce existential locations for simple references (which are only
used by desugaring), we use a new typing rule.
For technical reasons, we use a \verb+let+-expression to type check reference
allocation \emph{along with} a subsequent expression $e$ as a way to describe
the scope of $\locName$.

\equationSizeBegin
$$\inferrule*
  {\relTypVal{\Gamma}{\heapTyp}{\varVal}{\varTyp}{\maybeN} \sepPremise
   \relTypExp{\Gamma, \tyBind{x}{\utRef{\locName}}}
             {\heapCat{\heapTyp}{\heapBind{\locName}{\varVal}}}
             {e}{\varScm}{\heapTyp'}{\maybeN}}
  {\relTypExp{\Gamma}{\heapTyp}
             {\letinBare{x}{\newref{\locName}{\varVal}}{e}}
             {\existsLoc{\locName}{\varScm}}{\heapTyp'}{\maybeN}}
$$
\equationSizeEnd

\noindent
To facilitate algorithmic type checking, we ensure that existential locations
are always prenex quantified in types.
The \helperOp{Join} procedure, used for conditionals, rearranges existential
locations allocated on different branches to maintain this invariant.

Finally, we need to handle subtyping of existential location types.
The simplest approach is to require that two types have the same
quantifier structure.

\equationSizeBegin
$$\inferrule*
  {\relSub{\Gamma}{\varTyp_1}{\varTyp_2}{\maybeN}}
  {\relSub{\Gamma}
          {\existsLoc{\locName}{\varTyp_1}}
          {\existsLoc{\locName}{\varTyp_2}}{\maybeN}}
$$
\equationSizeEnd

\noindent
 For first-order functions, we can work around
this limitation by playing tricks with dummy locations. For higher-order
functions, however, the presence of existential locations limits expressivness
by constraining the use of the heap.
Abstracting over the mutable state of higher-order functions, however, can
can be quite heavyweight (see \eg Hoare Type Theory~\cite{HTT});
adding more lightweight support in our setting is left for future work.

\bibliographystyle{abbrv}
\bibliography{sw}

\end{document}